\documentstyle[prl,aps,epsf]{revtex}
\tighten
\begin{document}
\draft

\title{Incomplete Photonic Bandgap as Inferred from
the Speckle Pattern of Scattered Light Waves
%Disorder-Induced Intensity Correlations of Light Waves
}

\author{V. M. Apalkov$^1$, M. E. Raikh$^1$, and B. Shapiro$^2$ \\
$^1$Department of Physics, University of Utah, Salt Lake City,
UT 84112, USA \\
$^2$Department of Physics, Technion-Israel Institute of
Technology, Haifa 32000, Israel}
\maketitle

\begin{abstract}
Motivated by  recent experiments on intensity correlations of the
waves transmitted through disordered media, we demonstrate that the
speckle pattern from disordered photonic crystal with incomplete
band-gap represents a sensitive tool for determination the
stop-band width. We establish the  quantitative relation between this
width and the {\em angualar anisotropy} of the
intensity correlation function.
\end{abstract}

\noindent {\em Introduction.}
A wave propagating in a random medium undergoes multiple scattering
 and forms a complicated intensity pattern, commonly referred to as
 a specle pattern. It is described in statistical terms, with the
 help of probability distributions and correlation functions. The
 intensity correlation function, $\langle \delta I(\bbox{r})
 \delta I (\bbox{r}^{\prime}) \rangle $, where $\delta I(\bbox{r})$
 is the deviation from the averaged intensity at point $\bbox{r}$,
 contains a short range term\cite{shapiro86},
$C_1 ( \bbox{r}, \bbox{r}^{\prime})$,
 which oscillates on a scale of the wave length and
 exponentially decays beyond the mean free path, $l$.
 It also contains small long-range
terms\cite{stephen87,feng88,akkermans04},
which become  dominant
 for $| \bbox{r} - \bbox{r}^{\prime }| \gg l $.
Although
 the theory of speckle patterns was developed some time ago,
 it is only recently that the first experimental measurements
of the spatial correlator
 $C_1 ( \bbox{r}, \bbox{r}^{\prime})$ were
reported
 both for microwaves \cite{sebbah02} and optical
waves\cite{emiliani03,apostol}. These
 experiments were carried out for  {\em isotropic} disordered media;
macroscopic isotropy was also assumed in the existing theories.
In this isotropic situation comparison with the theory has enabled
the authors\cite{emiliani03} to infer the effective refractive index,
which
is the only relevant
%intrinsic
parameter of the  medium in the absence
of disorder.

The main message of
this paper is that, in a medium
with  {\em underlying  spatial structure},
the pattern of  intensity correlations exhibits a vastly richer
behavior as compared to the isotropic case.
Moreover, additional features in
$C_1(\bbox{r},\bbox{r}^{\prime})$ carry {\em quantitavive}
information
about this structure. As an example,
we consider
 a disordered incomplete-bandgap photonic crystal
and demonstrate how the band-structure parameters can be extracted
from the {\em angular anisotropy} of the correlator $C_1$.

Superficially, it may seem
that a wave with frequency $\omega$, arriving from
 a distant source, after having been scattered by many random
inhomogeneities,
 will lose all information about the crystal band-structure.
 Our point, though, is that  $C_1 ( \bbox{r}, \bbox{r}^{\prime})$ is
 essentially a {\em local} object: It is determined by scattering events
 in the vicinity of the points $\bbox{r}$, $\bbox{r}^{\prime}$, and it
is not sensitive to the ``prehistory'', i.e., scattering events
experienced by the wave before arrival to the point $\bbox{r}$.
In other words, $C_1 ( \bbox{r}, \bbox{r}^{\prime})$ can serve as
a "microscop" for observation
 of the local interference picture (on a scale smaller than $l$),
 thus, revealing the band-structure of the
inherent photonic crystal.
Such a microscop is particularly well suited for determination
of the band-structure of {\em realistic}
photonic crystals,
since it is not affected by the long-range disorder.
On the other hand, long-range disorder is  a generic feature of
realisitic crystals, such as silica-based synthetic opals.

\noindent {\em Disorder in synthetic opals.} Silica-based synthetic
opals, which are the fcc self-assemled arrangement of almost
monodispersed silica spheres, play a distinguished role in fabrication
of photonic band gap materials, the potential of which for light
manipulation was first appreciated in
Refs.\onlinecite{yablonovitch,john87}.  This is because the opals
constitute a template subsequently infiltrated with high refraction
index material, while the spheres are selectively removed (see {\em
e.g.}  \cite{wijnhoven98,zakhidov,blanco00}).  Due to this
distinguished role, the vast majority of experimental studies of light
propagation in photonic crystals has been carried out on
opals\cite{astratov02}.

In opal photonic crystals, the contrast of the dielectric constant
is weak, so that photonic bandgaps are not only
incomplete, but the corresponding stop-bands are relatively
narrow. For this reason, the photonic band structure of this materials
is strongly obscured by the disorder. As a result, any reliable
determination of the stop-band either from disorder-broadened
reflection maxima or from transmission minima, which are also srongly
affected by the disorder, is highly ambiguous. In fact, in early
studies, actual extraction  of the stop-band width from the data
was based on the ``thumb rules''\cite{vos96,tarhan96}.

Later
studies\cite{astratov02,vlasov99,vlasov00',baryshev03} have
yielded a deeper insight into the microscopic origin of the disorder
in opals. Namely, they indicated that one should distinguish three types
of the disorder

\noindent (i) Short-range disorder due to point defects and
spread in the sphere diameters. This type of the disorder causes
the mean free path\cite{huang01} $l\sim 15 \mu m$.
 
\noindent (ii) Stacking faults in $[111]$ direction\cite{vlasov99}.
This disorder has a {\em one-dimensional} character, and thus might
give rise to the 1D in-gap states\cite{deych98} within the incomplete
stop-band. In fact, experimentally observed deep and sharp
transmission minimum\cite{vlasov99} near the stop-band center
has been accounted for these states.
Subsequent interpretation of reflection
and transmission\cite{vlasov00'} as well as diffraction data
\cite{baryshev03} also relied on the prominent role played by the
stacking faults.

\noindent (iii) macroscopic domains that are $\sim
 50-100\mu m$ in size. Such domains inavoidably emerge in course
of self-assembly of thick enough photonic crystals\cite{astratov02}.
Due to their  presence
the stop-bands are strongly {\em
 inhomogeneously} broadened -- a serious complication for determination
 of photonic band structure.

Summarizing,
experimental
studies\cite{astratov02,vlasov99,vlasov00',baryshev03}
have led to significant improvement of the
understanding of the
light propagation in
strongly-disordered incomplete-bandgap photonic crystals.
However, no quantitative information
about the
stop-band width  could be inferred from these experiments.
Below we demonstrate that  this  width naturally emerges in
  the {\em angular dependence} of the intensity correlator $C_1$.

\noindent{\em Correlator $C_1$ in a disordered photonic crystal.}
The short-range correlator
$C_1 (\bbox{r},\bbox{r}^{\prime }) $, is obtained
by factorizing the product of four fields $\psi( \bbox{r})$
in the average $\langle I(\bbox{r})
I ( \bbox{r}^{\prime }) \rangle $. This leads\cite{stephen87} to
the relation
$C_1 (\bbox{r},\bbox{r}^{\prime }) = |C_{\psi }( \bbox{r},
  \bbox{r}^{\prime })|^2$,
where
\begin{equation}
\label{field}
 C_{\psi }( \bbox{r},\bbox{r}^{\prime })  =
 \frac{\langle \psi( \bbox{r}) \psi^{\star }(\bbox{r}^{\prime })\rangle  }
 {\langle I(\bbox{r}) \rangle^{1/2}   \langle
I ( \bbox{r}^{\prime }) \rangle^{1/2} }
\end{equation}
is the field correlation function. It follows from the Bethe-Salpeter
equation that
\begin{equation}
\label{c_psi}
 C_{\psi }( \bbox{r},\bbox{r}^{\prime })  = \frac{4\pi}{l}
   \int d\bbox{r}_1 \langle G( \bbox{r},\bbox{r}_1) \rangle
                    \langle G^{\ast}( \bbox{r}^{\prime },\bbox{r}_1) \rangle,
\end{equation}
where $\langle G( \bbox{r},\bbox{r}_1) \rangle$ is the disorder-averaged
Green's function.
The physical meaning of Eq. (\ref{c_psi}) is that
correlation between points $\bbox{r}$, $\bbox{r}^{\prime }$ is
established via scattering on an intermediate impurity, at $\bbox{r}_1$.
The integral Eq. (\ref{c_psi}) can be reduced to
\begin{equation}
\label{c_im}
 C_{\psi }( \bbox{r},\bbox{r}^{\prime })  = -
 \frac{4\pi c^3}{\epsilon ^{3/2}\omega }
    \mbox{Im} \langle G( \bbox{r},\bbox{r}^{\prime }) \rangle  ,
\end{equation}
which has the following intuitive explanation: field correlation
between a pair of points is due primarily to waves that are scattered
in the vicinity of one point and arrive to the vicinity of the
other point. The average amplitude to arrive from $\bbox{r}^{\prime }$
to $\bbox{r}$ is $\langle G( \bbox{r},\bbox{r}^{\prime }) \rangle$, so
that $C_{\psi }( \bbox{r},\bbox{r}^{\prime })  $ should be a linear
combination of $\langle G( \bbox{r},\bbox{r}^{\prime }) \rangle$ and
 $\langle G^{\star }( \bbox{r},\bbox{r}^{\prime }) \rangle$.
 Eq. (\ref{c_im}) is the right linear combination since it reduces
 to unity for  $\bbox{r}=\bbox{r}^{\prime }$. Using Eq. (\ref{c_im})
we present the field correlator as
\begin{equation}
\label{C_ImG}
 C_{\psi }(\bbox{r},\bbox{r}^{\prime }) = -
  \frac{4\pi c^3}{\epsilon ^{3/2}\omega }\mbox{Im}
\sum _{\mu, \bbox{k}} \frac{ \psi_{\mu, \bbox{k}}(\bbox{r})
    \psi^{\ast }_{\mu, \bbox{k}}(\bbox{r}^{\prime }) }{ \omega ^2
     - \omega _{\mu, \bbox{k}}^2 + i \eta } ,
\end{equation}
where $\eta =  c \omega /\epsilon^{1/2}l$, with $\epsilon$ standing for
the background dielectric constant, and $\psi^{\ast }_{\mu,
\bbox{k}}(\bbox{r})$,
%and
$\omega _{\mu, \bbox{k}}$ are the photonic crystal eigenmodes and
eigenfrequencies, respectively.
They are characterized by the wave vector $\bbox{k}$
of the first Brillouin zone and the band index $\mu = \pm 1$.

Suppose that stop-band corresponds to the direction
of propagation along the $z$-axis. Then, the anisotropy
of  $C_{\psi }(\bbox{r}, \bbox{r}^{\prime} )$ is expected within
a narrow angular interval,
$\theta = \arctan (z^{\prime }-z)/(\rho^{\prime }-\rho)\sim
\gamma ^{1/2}$, where $\gamma = \delta\omega/\omega \ll 1$ is the
dimensionless frequency width of the stop-band. For such $\theta$,
the main contribution to the
sum in Eq.~(\ref{C_ImG}) comes from the small domain of $\bbox{k}$-space
around the Bragg condition $k_z \approx Q_B/2$, where $2\pi/Q_B$
is the period along $z$ (see Fig. 1).
Within this domain the dispersion law, $\omega _{\mu, \bbox{k}}$,
can be simplified as
\begin{equation}
\label{energy}
\frac{\epsilon\omega _{\mu, \bbox{k}}^2  }{c^2} =
k_{\perp }^2 + \frac{Q_B^2}{4} + \mu
Q_B \sqrt{ \frac{1}{L_B^{2}} + \left( k_z-\frac{Q_B}{2}\right)^2  },
\end{equation}
where we have introduced the Bragg length,
$L_B=4\omega/Q_B\delta\omega$, of the
 decay along $z$  exactly at the Bragg condition.
 The corresponding eigenmodes, $\psi _{\mu , \bbox{k}}$
can be also simplified
\begin{equation}
\psi  _{\mu , \bbox{k}} = {\cal U}_{\mu } (z)
\exp\left[ i(k_z-Q_B/2)z + i\bbox{k}_{\perp} \bbox{\rho}\right],
\end{equation}
where ${\cal U}_{\mu }$ are the Bloch functions.
It is seen from Eq. (\ref{energy}) that, for a given
$\omega_{\mu , \bbox{k}}=\omega$ and the same $k_z$, the
values of $k_{\perp}$ are very different for upper $\mu=1$
and lower $\mu=-1$ bands. As a result, the contribution to $C_{\psi }$
from the upper band is exponentially smaller.
Then, keeping only $\mu =-1$ terms,
we can express the field-field correlation function as
%\begin{eqnarray}
%\label{C_psi}
% C_{\psi }(\bbox{r}, \bbox{r}^{\prime} )
%  & = &  -\frac{c^3 {\cal U}_{-1 }(z){\cal U}_{-1 }(z^{\prime }) }
% {2 \pi ^3 \epsilon^{3/2} \omega }
%\int dk_z
%  \int k_{\perp } d k_{\perp}\int d \phi
%\times \nonumber \\
%  &  &  \mbox{Im} \left\{
%  \frac{ \exp \left[i z (k_z- Q_B/2 )+i \rho k_{\perp }
%\cos \phi \right] }
%         { \omega ^2 - \omega _{-1,\bbox{k}}^2 + i \eta }
%\right\} .
%\end{eqnarray}
\begin{equation}
\label{C_psi}
 C_{\psi }(\bbox{r}, \bbox{r}^{\prime} )
  =  -\frac{c^3 {\cal U}_{-1 }(z){\cal U}_{-1 }(z^{\prime }) }
 {2 \pi ^3 \epsilon^{3/2} \omega }
\int dk_z
  \int k_{\perp } d k_{\perp}\int d \phi ~
 \mbox{Im} \left\{
  \frac{ \exp \left[i z (k_z- Q_B/2 )+i \rho k_{\perp }
\cos \phi \right] }
         { \omega ^2 - \omega _{-1,\bbox{k}}^2 + i \eta }
\right\} .
\end{equation}
Substituting
Eq.~(\ref{energy}) into  the integral
Eq.~(\ref{C_psi}), and performing the integration over
$k_{\perp }$ and $\phi $ we arrive at
%\begin{eqnarray}
% & & C_{\psi }(\bbox{r}, \bbox{r}^{\prime} )  =
%\left(\frac{\gamma}{4} \right)
%{\cal U}_{-1}(z) {\cal U}_{-1}(z^{\prime })   \times
%                                                \nonumber  \\
%& & ~~ \mbox{Re} \left\{
%\int dq
%\exp\left[ iq \frac{\gamma {\cal R}}{4}  \right]
%H_0 \left( \frac{ \theta
%{\cal R} \gamma ^{1/2} }{2 }
%  \sqrt{\Delta +\sqrt{1+q^2}}   \right)
%          \right\} ,
%\label{C_psi_2}
%\end{eqnarray}
\begin{equation}
 C_{\psi }(\bbox{r}, \bbox{r}^{\prime} )  =
\left(\frac{\gamma}{4} \right)
{\cal U}_{-1}(z) {\cal U}_{-1}(z^{\prime })
                                             \mbox{Re} \left\{
\int dq
\exp\left[ iq \frac{\gamma {\cal R}}{4}  \right]
H_0 \left( \frac{ \theta
{\cal R} \gamma ^{1/2} }{2 }
  \sqrt{\Delta +\sqrt{1+q^2}}   \right)
          \right\} ,
\label{C_psi_2}
\end{equation}
where $H_0$ is a Hankel function of a zero order.
The dimensionless length ${\cal R}$ and
``detuning'' $\Delta$ are defined as
\begin{equation}
\label{Delta}
{\cal R}=|\bbox{r}- \bbox{r}^{\prime}| Q_B, ~~~
\Delta =
\frac{2}{\delta \omega }
\left( \omega - \frac{cQ_B}{2\epsilon ^{1/2}} \right)
 + i \frac{L_B}{2 l}.
\end{equation}
For $\theta {\cal R}\gamma ^{1/2}\gg 1$, i.e.
$|\bbox{r}- \bbox{r}^{\prime}| \gg L_B$,
the asymptotic expansion of the
Hankel function can be used, so that the integral Eq.~(\ref{C_psi_2})
takes the form
%\begin{eqnarray}
%\label{C_psi_3}
% & & C_{\psi }(\bbox{r}, \bbox{r}^{\prime} )  =
%    {\cal U}_{-1}(z) {\cal U}_{-1}(z^{\prime })
%\left[\frac{ \gamma ^{3/4}}
%  {2 (\pi {\cal R})^{1/2}} \right] \times  \nonumber  \\
%& & ~~ \mbox{Re} \left\{ e^{-i\pi/4}
%\int dq \frac{\exp \left[ i {\cal R} F(\theta, q)\right] }
%{  \left( \Delta +\sqrt{1+q^2} \right)^{1/4}}
%          \right\} ,
%\end{eqnarray}
\begin{equation}
\label{C_psi_3}
 C_{\psi }(\bbox{r}, \bbox{r}^{\prime} )  =
    {\cal U}_{-1}(z) {\cal U}_{-1}(z^{\prime })
\left[\frac{ \gamma ^{3/4}}
  {2 (\pi {\cal R})^{1/2}} \right] ~ \mbox{Re} \left\{ e^{-i\pi/4}
\int dq \frac{\exp \left[ i {\cal R} F(\theta, q)\right] }
{  \left( \Delta +\sqrt{1+q^2} \right)^{1/4}}
          \right\} ,
\end{equation}
where $F(\theta, q ) = \gamma q/4 + \theta(\gamma^{1/2}/2)
\left[\Delta +\sqrt{1+q^2}\right]^{1/2}$.
Function $F(\theta, q)$ has a special point at which
the first and the second derivatives with respect to
$Q$, for some value of $\theta =\theta _c$, are zero.
The position of this critical point, $(\theta _c, q_c)$,
depends on the detuning $\Delta $ and  can be parametrized by a
complex number $w$ as
\[
\Delta = \frac{4}{27}w^3 - w, ~\theta_c = 2 \gamma^{1/2}\left(
\frac{w }{3}\right)^{3/2},
 ~q_c = \frac{1}{3}\left( 4 w^2 - 9 \right)^{1/2}
\]
For $\Delta = 0$ we have $w = 3^{3/2}/2$,
$\theta _c = (3^{3/4}/2^{1/2}) \gamma^{1/2}$,
and $q_c = 2^{1/2}$. To avaluate the integral
 Eq.~(\ref{C_psi_3}) for $\theta $ close to the critical point,
 $\theta \sim \theta _c $,  the function $F(\theta ,q)$
should be expanded up to the third order in $(q-q_c)$
\begin{equation}
\label{expansion}
F(\theta, q ) =  \phi_0 + \frac{\delta \theta}{\theta_t} +
\delta q\frac{\delta \theta }{\theta _d}   - \frac{1}{3} (\delta q)^3,
\end{equation}
where $\delta \theta = \theta - \theta _c$, $\delta q =
(9/4w)(6/w)^{1/3}(q - q_c )$,
$F_0=\gamma \left(4 w^2 - 9 \right)^{3/2}/108$, and characteristic
angles $\theta_t$ and $\theta_d$ are defined as
\begin{equation}
  \theta_t  =
\frac{6}{\gamma ^{1/2}}\left(\frac{3 }{4w^3-9w}\right)^{1/2},~~
  \theta _d = 6\left(\frac{2w}{3\gamma }\right)^{1/6} .
\label{constant}
\end{equation}
Substituting Eq.~(\ref{expansion}) into Eq.~(\ref{C_psi_3}) and
performing integration over $\delta q$, we obtain
%\begin{eqnarray}
%&  & C_{\psi }(\bbox{r},\bbox{r}^{\prime})  =
%\left[ \frac{ 2\gamma^{1/5}}{3{\cal R}}\right]^{5/6}
%{\cal U}_{-1}(z) {\cal U}_{-1}(z^{\prime})
%\mbox{Re} \left\{
%   \frac{\pi^{1/2} w^{1/3}}{(4w^2-9)^{1/4}} \right.
%\nonumber \\
% & &  \times \left. \exp \left(
%i \phi_0
% +i{\cal R}\frac{\delta \theta }
%   {\theta_t} \right) \mbox{\em Ai} \left(- {\cal R}^{2/3}\frac{\delta
%                              \theta }{\theta_d}
%    \right)
%  \right\}  ,
%\label{final}
%\end{eqnarray}
\begin{equation}
C_{\psi }(\bbox{r},\bbox{r}^{\prime})  =
\left[ \frac{ 2\gamma^{1/5}}{3{\cal R}}\right]^{5/6}
{\cal U}_{-1}(z) {\cal U}_{-1}(z^{\prime})
\mbox{Re} \left\{
   \frac{\pi^{1/2} w^{1/3}}{(4w^2-9)^{1/4}}  \exp \left(
i \phi_0
 +i{\cal R}\frac{\delta \theta }
   {\theta_t} \right) \mbox{\em Ai} \left(- {\cal R}^{2/3}\frac{\delta
                              \theta }{\theta_d}
    \right)
  \right\}  ,
\label{final}
\end{equation}
where $\mbox{\em Ai} $ is the Airy function.

\noindent{\em Discussion}. It is easy to see that Eq. (\ref{final})
has a form similar to the field distribution near the caustics, which
is the envelope of rays reflected  by a curved surface\cite{landau80}.
This similarity is not accidental. Indeed, near the caustics the
light rays cross over from geometrically allowed directions
to the ``shadow''. Eq. (\ref{final}) also describes a crossover
 from the directions  $\theta \gg \theta_c\sim \gamma^{1/2}$
of free propagation
to the angular domain $\theta \lesssim \gamma^{1/2}$
where propagation is forbidden due to the stop-band, which plays the
role of ``shadow'' in the $\bbox{k}$-space.
Similarly to the case of caustics, the behavior of $C_{\psi}$
 near the  crossover
$\theta =\theta_c$
is described by the following exponential decay at $\theta < \theta_c$
\[
C_{\psi } (\theta ) \sim \exp \left[- {\cal R} \left(\frac{\theta _c -
\theta }
{\theta _d}  \right)^{3/2}   \right] \cos\left[ {\cal R} \frac{\theta_c -
\theta }
{\theta _t}+\phi _0  \right],
\]
which, in the case of intensity correlation serves as an envelope
of the fast oscillations. For $\theta > \theta_c$ this envelope
oscillates itself as $\cos \left[ {\cal R}
\left(\frac{\theta
 - \theta _c} {\theta _d}  \right)^{3/2}  - \pi/4  \right]$. Overall,
there are three angular scales in the correlator $C_{\psi}$,
namely, the critical angle, $\theta_c$, the period of the envelope,
$\theta_d/{\cal R}^{2/3}$, and the period, $\theta_t/{\cal R}$,
of the oscillations. At spatial distances
$|\bbox{r} -\bbox{r}^{\prime }|\sim L_B$ all three scales are of
the same order $\sim \gamma^{1/2}$. For large
$|\bbox{r} -\bbox{r}^{\prime }|$ we have $\theta_c \gg
\theta_d/{\cal R}^{2/3} \gg \theta_t/{\cal R}$. Numerical examples,
illustrating separation of the three scales is given in Fig. 2.
Outside the stop-band, $\theta \gg \gamma^{1/2}$ the correlator $C_1$
saturates at the  value
$\propto \exp\left(-|\bbox{r} -\bbox{r}^{\prime }|/l\right)$,
governed by the mean free path\cite{shapiro86}. Obviously, our
main result Eq. (\ref{final}) applies in the vicinity of each of
the equivalent Bragg directions.

When large-scale disorder (domains) broadens significantly
the reflectivity peak, $\bbox{R}(\theta)$, the dependence
$C_1(\theta)$ with all fine details {\em remains unchanged}.
Even if disorder is purely short-range, the
intensity correlations are much more sensitive to the
periodic background than the reflectivity. To illustrate this,
in Fig. 3 we present the dependence $C_1(\theta)$ together with
$\bbox{R}(\theta)$, which near the Bragg condition is given by
\begin{equation}
\bbox{R} = \left|  \frac{
           \epsilon ^{1/2}-1 + \left(\epsilon ^{1/2}+1\right)
  \left[ g(\theta ) - \sqrt{ 1+ g^2(\theta ) }  \right]
                        }{
          \epsilon ^{1/2}+1 + \left(\epsilon ^{1/2}-1\right)
  \left[ g(\theta ) - \sqrt{ 1+ g^2(\theta ) }  \right]}
  \right|^2,
\end{equation}
where $g(\theta )= L_B/2l+4i\theta^2/\epsilon \gamma$.
The dependencies $\bbox{R}(\theta)$, $C_1(\theta)$ were calculated
for the  effective refractive
index $\epsilon^{1/2}=1.7$ and $L_B=l$. Similar to $C_1$,
the width of $\bbox{R}(\theta)$ is $\theta \sim \gamma^{1/2}$
However, $\bbox{R}(\theta)$ is strongly smeared when $L_B=l$,
whereas $C_1(\theta)$ contains oscillations. Three
 oscillations are well pronounced for
the distances $|\bbox{r}-\bbox{r}^{\prime}|$ equal to $2l$,
and $3l$, as shown in Fig. 3.
Since these values are comparable to $L_B$,  the argument of
the Hankel function in Eq. (\ref{C_psi_2}) is $\sim 1$. Therefore,
$C_1=\vert C_{\psi}\vert^2$ was calculated directly from
(\ref{C_psi_2}) without using the asymptotical form
Eq. (\ref{C_psi_3}).
Therefore, even when $L_B \sim l$,
comparison of the saturation values of $C_1(\theta)$ at two
$|\bbox{r}-\bbox{r}^{\prime}|$ allows to determine the value of
the mean free path. Then $L_B$ and, correspondingly, the stop-band
width, $\delta \omega = 2c/\epsilon^{1/2}L_B$, can be inferred from
the oscillations in the following way.  Knowledge of the
oscillation period,
$\pi\theta_t(w)/{\cal R}$ together with frequency and mean free path,
allows to deduce the parameter $w$, then
the detunig $\Delta$, defined by Eq.~(\ref{Delta}), and, finally,
$L_B=2l~\!\bigl\{\mbox{Im} \Delta\bigr\} $.

\noindent {\em Conclusion}. In the present paper we
have demonstrated that the intensity correlations in a
disordered photonic crystal with {\em incomplete} bandgap
uncover the underlying band structure that is obscured by the disorder,
so that conventional approaches applicable to the ``clean'' photonic
crystals\cite{galisteo03,thijssen99} fail.

It is important
to emphasize that the very phenomenon of intensity correlation is
 a {\em disorder induced} effect and that the
function $C_1 (\bbox{r}, \bbox{r}^{\prime })$ is calculated under
the condition of diffusive propagation of the wave inside the sample.
Note that, conceptually, with $C_1$ playing the role
of a microscope, there
is an analogy between the correlation of the fluctuations and the
experimental setup\cite{petrov98,romanov02,koenderink02},
in which the source of light is located inside the crystal and the
directionality of the outgoing light is studied.
On the contrary, the long-range contributions\cite{stephen87,feng88}
to the intensity correlator come from scattering trajectories
that explore {\em all} directions inside the crystal, and, due
to isotropization, do not carry information about the Bragg directions.
 
Note in conclusion, that sensitivity of a number of disorder-induced
effects, such as coherent backscattering \cite{huang01,sivachenko01}
and directionality of the exiting diffusing light\cite{koenderink03},
to the periodic background has already been discussed in the
literature.  This sensitivity has also been employed
\cite{huang01,koenderink03} for the estimation of the stop-band
width. In this regard, the intensity correlations constitute a much
more accurate tool. Indeed, in the shape of the coherent
backscattering cone, the properties of periodic structure enter only
through the convolution with cooperon, which describes the large-scale
diffusive motion, whereas the correlator $C_1$ comes from {\em short
distances}.  Another advantage of using intensity correlations over
traditional methods is that $C_1$ is insensitive to the long-range
disorder due to domains that mask the photonic band
structure\cite{astratov02}. Earlier it was demonstrated
\cite{vlasov00} that, isolating a single domain and measuring
reflection from this domain, results in a drastic narrowing of the
reflection spectra. We emphasize, that correlation analysis of the
speckle pattern allows to get rid of inhomogeneous broadening in a
natural way.

Finally, an attractive avenue for the future work comes
from proposed infiltrating opals
with nematic liquid crystals
 to manupulate the photonic band gap externally\cite{busch99}.
While in reflectivity measurements reorientation of the
nematic director manifests itself in a simple shift of the optical
Bragg reflection peak\cite{kang01}, the structure of the
correlator $C_1$ might reveal  the effect of birefringence on the
photonic bandstructure.

\noindent{\em Acknowledgements}. One of the authors (B.S.) acknowledges
the hospitality of the University of Utah. The work was supported
by the Army Research Office under grant No. DAAD 19-03-1-0290 and
by the Petroleum Research Fund under grant No. 37890-AC6.

%\end{document}

\begin{figure}
\centerline{
\epsfxsize=2.7in
\epsfbox{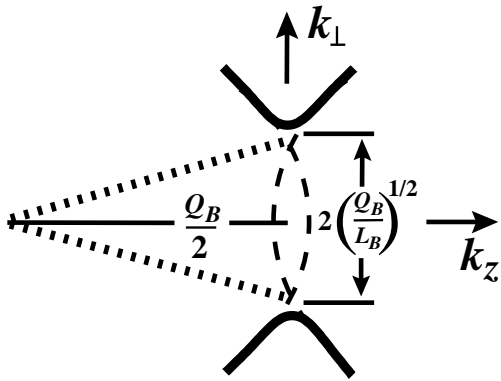}
}
\vspace*{0.1in}
\protect\caption[sample]
{\sloppy{Surface of constant frequency $\omega(k_z,k_{\perp})=\omega$
is shown schematically for the Bragg frequency
$\omega=cQ_B/2\epsilon^{1/2}$}}
\label{figone}
\end{figure}

\begin{figure}
\centerline{
\epsfxsize=3.3in
\epsfbox{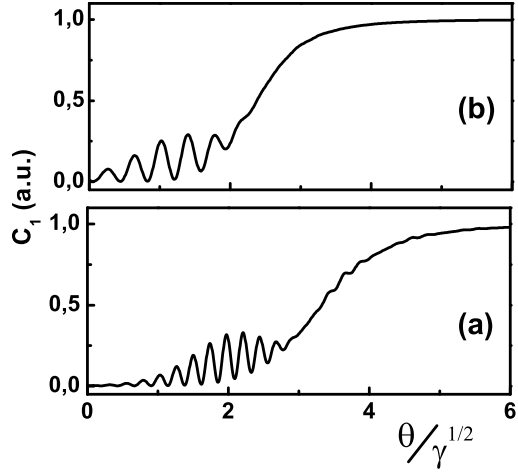}
}
\vspace*{0.1in}
\protect\caption[sample]
{\sloppy{Correlator $C_1$ as a function of dimensionless angle
$\theta/2\left(Q_BL_B\right)^{-1/2}$ is shown for
$|\bbox{r}-\bbox{r}^{\prime}|=5L_B$ and for two dimensionless
``detunings'', $\Delta$, defined by Eq. (\ref{Delta});
(a) $\Delta = 0.15i$  (i.e., frequency is equal to the Bragg frequency
and $L_B = 0.3l$); (b) $\Delta =0.5 + 0.15i$. }}
\label{figtwo}
\end{figure}

\begin{figure}
\centerline{
\epsfxsize=3.3in
\epsfbox{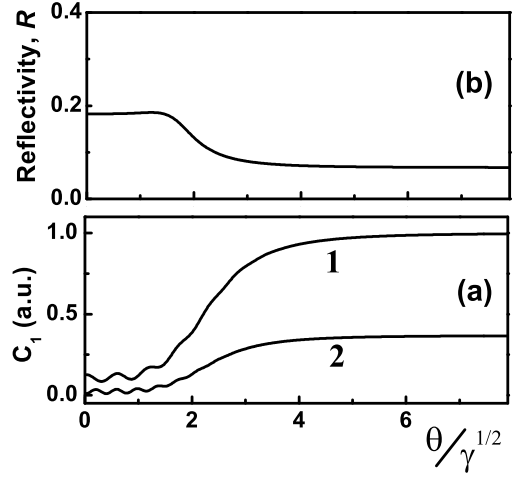}
}
\vspace*{0.1in}
\protect\caption[sample]
{\sloppy{(a) Correlator $C_1$ as a function of dimensionless angle
$\theta/2\left(Q_BL_B\right)^{-1/2}$ is shown for $L_B=l$ and two
distances
$|\bbox{r}-\bbox{r}^{\prime}|=2l$ (curve 1) and
$|\bbox{r}-\bbox{r}^{\prime}|=3l$ (curve 2); (b) Angular dependence
of the reflectivity is shown for  $L_B=l$ and effective refractive
index $\epsilon^{1/2}=1.7$.
}}
\label{figthree}
\end{figure}

\end{document}